\documentclass[useAMS,usenatbib]{mn2e}

\usepackage{graphicx}
\usepackage{times}
\usepackage{natbib}
\usepackage{amsmath}

\renewcommand{\d}{\mathrm{d}}
\newcommand{\gsim}{\ga}

\newcommand{\gtrsim}{\ga}
\newcommand{\lesssim}{\la}

\def\Zcrit{{\rm Z_{crit}}}
\def\zsun{{\rm Z_\odot}}
\def\msun{{\rm M_\odot}}
\def\msunh{{\rm M_\odot/{\it h}}}
\def\fnl{$f_{\rm NL}$}
\def\Mpch{{\rm Mpc/{\it h}}}

\def\Omegab{{\Omega_{0,\rm b}}}
\def\Omegam{{\Omega_{0,\rm m}}}
\def\Omegal{{\Omega_{0,\rm \Lambda}}}

\title[LGRBs as probes of Non-Gaussianities]{Counts of high-redshift GRBs as probe of primordial non-Gaussianities}
\author[U.~Maio et al.]{
  Umberto~Maio$^1$\thanks{E-mail: umaio@mpe.mpg.de},
  Ruben Salvaterra$^2$,
  Lauro Moscardini$^{3,4,5}$, and
  Benedetta Ciardi$^6$
  \\
  ${}^1$Max-Planck-Institut f\"ur extraterrestrische Physik, Giessenbachstra{\ss}e 1,  D-85748 Garching b. M\"unchen, Germany\\
  ${}^2$ INAF/IASF - Italian National Astrophysical Institute, via E. Bassini 15, I-20133 Milano, Italy\\
  ${}^3$ Dipartimento di Astronomia, Universit\`a di Bologna, via Ranzani 1, I-40127 Bologna, Italy\\
  ${}^4$ INAF, Osservatorio Astronomico di Bologna, via Ranzani 1, I-40127 Bologna, Italy\\
  ${}^5$ INFN, Sezione di Bologna, viale Berti Pichat 6/2, I-40127 Bologna, Italy\\
  ${}^6$ Max-Planck-Institut f\"ur Astrophysik, Karl-Schwarzshild-Stra{\ss}e 1,  D-85748 Garching b. M\"unchen, Germany
}

\begin{document}

\date{Accepted 2012 August 1. Received 2012 July 31; in original form 2012 July 17}

\pagerange{\pageref{firstpage}--\pageref{lastpage}}\pubyear{0}
\maketitle
\label{firstpage}

\begin{abstract}
  We propose to use high-redshift long $\gamma$-ray bursts (GRBs) as
  cosmological tools to constrain the amount of primordial non-Gaussianity 
  in the density field. By using numerical, N-body, hydrodynamic,
  chemistry simulations of different cosmological volumes with various
  Gaussian and non-Gaussian models, we self-consistently relate the
  cosmic star formation rate density to the corresponding GRB rate.
  Assuming that GRBs are fair tracers of cosmic star formation, we
  find that positive local non-Gaussianities, described in terms of the
  non-linear parameter, \fnl, might boost significantly the GRB rate at
  high redshift, $z \gg 6$. Deviations with respect to the Gaussian
  case account for a few orders of magnitude if \fnl$\sim 1000$, one
  order of magnitude for \fnl$\sim 100$, and a factor of $\sim 2$ for
  \fnl$\sim 50$.  These differences are found only at large redshift,
  while at later times the rates tend to converge.
  Furthermore, a comparison between our predictions and the observed
  GRB data at $z > 6$ allows to exclude large negative \fnl, consistently 
  with previous works. Future detections of any long GRB
  at extremely high redshift ($z\sim 15-20$) could favor non-Gaussian
  scenarios with positive \fnl.
  More stringent constraints require much larger high-$z$ GRB complete 
  samples, currently not available in literature.
  By distinguishing the contributions to the GRB rate from the 
  metal-poor population~III regime, and the metal-enriched population~II-I 
  regime, we conclude that the latter is a more solid tracer of the underlying 
  matter distribution, while the former is strongly dominated by feedback
  mechanisms from the first, massive, short-lived stars, rather than
  by possible non-Gaussian fluctuations. This holds quite independently
  of the assumed population~III initial mass function.
\end{abstract}



\section{Introduction}\label{Sect:introduction}


The present standard cosmological model assumes that a primordial
inflationary phase \cite[][]{Starobinsky1980,Guth1981,Linde1990} ends
with the creation of density fluctuations, that then grow during
cosmological times \cite[e.g.][]{GunnGott1972, Weinberg1972,
   PressSchechter1974, WhiteRees1978, Peebles1993, ShethTormen1999,
   Peacock1999, Hogg1999astro.ph, ColesLucchin2002, PR2003} to give
birth to the presently observed large scale structure of the Universe
\cite[][]{BarkanaLoeb2001, CiardiFerrara2005, BrommYoshida2011}.
Stars, galaxies, and clusters of galaxies form by gravitational
collapse in an expanding flat space, composed by $\sim 30\%$ of matter
and $\sim 70\%$ of an unknown constituent referred to as dark
energy, for which the cosmological constant $\Lambda$ represents the
simplest explanation. Thanks to the evidences coming from different
observational datasets (mainly cosmic microwave background, galaxy
surveys and supernovae), the general properties of our Universe have become
clearer and its parameters known with much better accuracy. 
The estimated contributions to the cosmic density are \cite[][]{Komatsu2011}:
$\Omegam = 0.272$, $\Omegal = 0.728$, $\Omegab = 0.044$, 
for matter, 
cosmological constant, 
and baryons, respectively; 
the cosmic equation of state parameter is consistent with $w=-1$, 
the theoretical expectation of the cosmological constant; 
the primordial power spectrum has spectral index $n=0.96$, and 
a normalization corresponding to a mass variance within a 
$8~\rm\Mpch$-sphere of $\sigma_8=0.8$.
\\
\noindent
Even if the above picture is quite satisfying, the specific
mechanism driving the inflation is however not completely
understood. This fact justifies the existence in the literature of a
plethora of possible inflationary models, each of them with specific
predictions for various observables. In particular, the study of the
statistical distribution of the primordial fluctuations is considered
one of the best way to discriminate between them.
In fact, alternatives to the standard single-field slow-roll model, which 
predicts a nearly Gaussian distribution, can produce a significant amount of
non-Gaussianity \cite[][]{Bartolo2004, Chen2010}.
The most recent analyses of the observational data show some evidence for 
possible departures from Gaussianity, even if with a low level of 
significance (\citealt{Peebles1983, DesjacquesSeljak2010, LoVerde2011arXiv, DAmico2011, Komatsu2011}; as
also collected in the summary Table~2 by \citealt{MaioIannuzzi2011}).
More precisely, slightly positively skewed models are favored.
\\
From a theoretical point of view, the presence of some amount of
primordial non-Gaussianity has two main effects, which are then used
as efficient constraining tools:
it introduces a scale-dependence in the bias factor 
\cite[e.g.][]{Grinstein1986, McDonald2008, Desjacques2009, Fedeli2011, Norena2012arXiv, Wagner2012}, 
and it modifies the abundance and the formation history of rare events
(i.e. very low- and high-sigma fluctuations; e.g. \citealt{Koyama1999, Zaldarriaga2000, Grossi2007, Grossi2008, Wagner2010, LoVerde2011arXiv}).
High redshifts represent an interesting regime to potentially investigate
these effects.
Indeed, very early structures and primordial mini-haloes hosting the 
first bursts of star formation are expected to be somehow affected by 
the presence of primordial non-Gaussianities
\cite[as discussed in][]{Maio2011cqg}.
\\
In more detail, due to the sensitivity of the gas cooling capabilities to
the underlying matter density field, numerical hydrodynamical 
simulations have shown that the initially skewed non-Gaussian features 
could be reflected
by  the probability distribution function of the high-$z$ cosmic medium 
\cite[][]{Viel2009},
by a change in the molecular gas evolution and formation epoch of
first stars and galaxies \cite[][]{MaioIannuzzi2011, Maio2011cqg},
and by the consequent metal pollution in the Universe 
\cite[][]{MaioKhochfar2012}.
Furthermore, simple semi-analytical arguments have suggested 
non-Gaussianity effects on the birth of primordial black holes 
\cite[e.g.][]{BullockPrimack1997, GreenLiddle1997, Ivanov1998, Avelino2005, Hidalgo2007arXiv, Kohri2008, BugaevKlimai2012, Byrnes2012arXiv},
cosmic reionization \cite[][]{Crociani2009},
and hydrogen 21-cm signal \cite[][]{Cooray2006, Cooray2008, Pillepich2007, Joudaki2011, ChongchitnanSilk2012arXiv}.
\\
In this respect, a key tool for studies of high-redshift environments
might be represented by $\gamma$-ray bursts (GRBs), powerful explosions
emitting $\gamma$ rays in the $\rm\sim [1~keV, 10~MeV]$ energy band,
mostly distributed around $\sim 0.1-1\,\rm MeV$,
as detected by the latest {\it Fermi-GBM} instrument operating in the 
[8~keV, 40~MeV] range \citep{Bissaldi2011Fermi}.
\\
These bursts have:
isotropic equivalent peak luminosities as high as $\sim 10^{54}$ erg s$^{-1}$
\cite[the record holder being GRB~080607, ][]{Perley2011};
an isotropic angular distribution
\cite[][]{Fishman1994, Paciesas1999, Paciesas2012Fermi};
and a bimodal duration distribution \cite[][]{Kouveliotou1993},
with most of them lasting for a period longer than 2~seconds (long GRBs),
and  some of them, detected mostly at low redshift, for a period shorter 
than 2~seconds (short GRBs).
\\
In the following we will only consider long GRBs (LGRBs), which are 
supposedly related to the death of massive stars
\cite[see extensive reviews by e.g.][]{Piran2004, Meszaros2006} and, 
therefore, they are indicators of the local star formation episodes
\cite[e.g.][]{Jakobsson2005, Nuza2007, Lapi2008, 
Yuksel2008, Kistler2009, Butler2010, Campisi2011b, Mannucci2011, Ishida2011, Elliott2012, RobertsonEllis2012, Michalowski2012arXiv}.
\\
Typical $\gamma$-ray bursts have long-lasting afterglows
at lower frequencies, from the X-rays to the radio band
due to scattering with the surrounding ambient medium
\cite[][]{Paczynski1991, Dermer1992},
and are theoretically explained by the ``collapsar model''
\cite[][]{Woosley1993, WangWeeler1998, Meszaros1999, WoosleyHeger2012}:
a massive black-hole stellar remnant -- probably a Wolf-Rayet star
\cite[but see][]{Baron1992, Yoonetal2010} -- accreting stellar mass 
from a disk
\cite[][]{Popham1999, Fryer1999, Narayan2001, YoonLanger2005,DeColle2012}
at a rate of $\sim 0.01-10\,\rm\msun/s$,
and accompanied by a collimated-jet emission with a few degree opening angle
\cite[e.g.][]{Waxman1997a, Rhoads1997ApJ, Sari1998, WangWeeler1998,
 Schmidt1999, Schmidt2001}.
\\
Due to additional factors, like asymmetric explosions or stellar rotation
\cite[e.g.][]{Sollerman2005, Thone2008, Ostlin2008},
only a small fraction of SNe, $\sim 10^{-2}-10^{-3}$
\cite[][]{Fruchter2006, Yoon2006, Bissaldi2007, Soderberg2010, 
Grieco2012arXiv}, can result into a LGRB.
However, also taking into account such effects, there is still significant 
lack of knowledge of some important details, like the minimum mass for GRB
black-hole progenitors, that is highly debated and expected to lie between
typical SN limits of $\sim 25 - 40\,\rm M_\odot$
\cite[see also recently proposed upper values of even $\sim 60\,\rm M_\odot$ in][]{Georgy2012}.
\\
The uniquely bright luminosities of GRBs facilitate their detection up
to very high redshift, as shown by the three bursts spectroscopically
confirmed at $z>6$, i.e.
GRB~050904 at $z=6.3$ \cite[][]{Kawai2006},
GRB~080913 at $z=6.7$ \cite[][]{Greiner2009}, and 
GRB~090423 at $z=8.2$ \cite[][]{Salvaterra2009, Tanvir2009}, 
and by the case of GRB~090429B, having a photometric redshift of
$z\sim 9.4$ \cite[][]{Cucchiara2011}.
\\
High-redshift GRBs are a powerful and, in some cases, a unique tool to
study the Universe at the early stages of structure formation and can
provide fundamental information about the environment of their own
{\it hosting galaxies} like:
\begin{itemize}
\item metallicity and dust content \cite[][]{Savaglio2005,
    Savaglio2006, Nuza2007, Fynbo2008, Savaglio2009, Mannucci2011,
    Campisi2011b, Niino2011};
\item neutral-hydrogen fraction \cite[][]{Nagamine2008, McQuinn2008,
    McQuinn2009, Gallerani2008, Miralbel2011, RobertsonEllis2012};
\item
  local inter-galactic radiation field \cite[][]{Inoue2010};
\item
  early cosmic magnetic fields \cite[][]{Takahashi2011};
\item
  stellar populations \cite[][]{BrommLoeb2006, Campisi2011, deSouza2011, Salvaterra2012};
\end{itemize}

\noindent
In the present work, we argue that GRBs can be additionally used as
cosmological probe of the amount of non-Gaussianity present in the
primordial density field. In fact, they are sensitive to the underlying
cosmological model through the first episodes of the cosmic star
formation history.
\\
We will show how GRBs trace the matter distribution at high redshift by
performing a detailed analysis of the GRB rate in different non-Gaussian
scenarios, with the help of N-body, hydrodynamic, chemistry simulations 
of early structure formation \cite[][]{MaioIannuzzi2011}.
In the simulated volumes, star formation is addressed on the basis of the
local thermodynamical properties of the collapsing gas, by
consistently following its density, temperature and chemical
composition, and by taking into account stellar evolution and feedback
effects.
The resulting star formation rate (SFR) and the adopted
initial mass function (IMF) for the stellar populations tracked during
the runs are used to determine the expected GRB formation rate 
density in the various cases, 
and hence the integrated GRB rate ($R$), for both
metal-poor population~III (hereafter, popIII) regime and
metal-enriched population~II-I (hereafter, popII-I) regime.
\\
The paper is structured as follows.
In Sect.~\ref{Sect:simulations} we describe the numerical simulations 
used in our study;
in Sect.~\ref{Sect:results} we present the adopted model for GRB
evolution (Sect.~\ref{sect:model}),
its validation (Sect.~\ref{sect:swiftvalidation}), 
and the consequencies for non-Gaussian models (Sect.~\ref{Sect:FinalResults});
finally, in Sect.~\ref{Sect:discussion} we discuss and summarize our 
findings and conclude.
In the following, when mentioning GRBs we will refer to LGRBs.

\label{tab:parameters} 


\section{Numerical Simulations}\label{Sect:simulations}


\begin{figure*}
\centering
\includegraphics[width=0.49\textwidth]{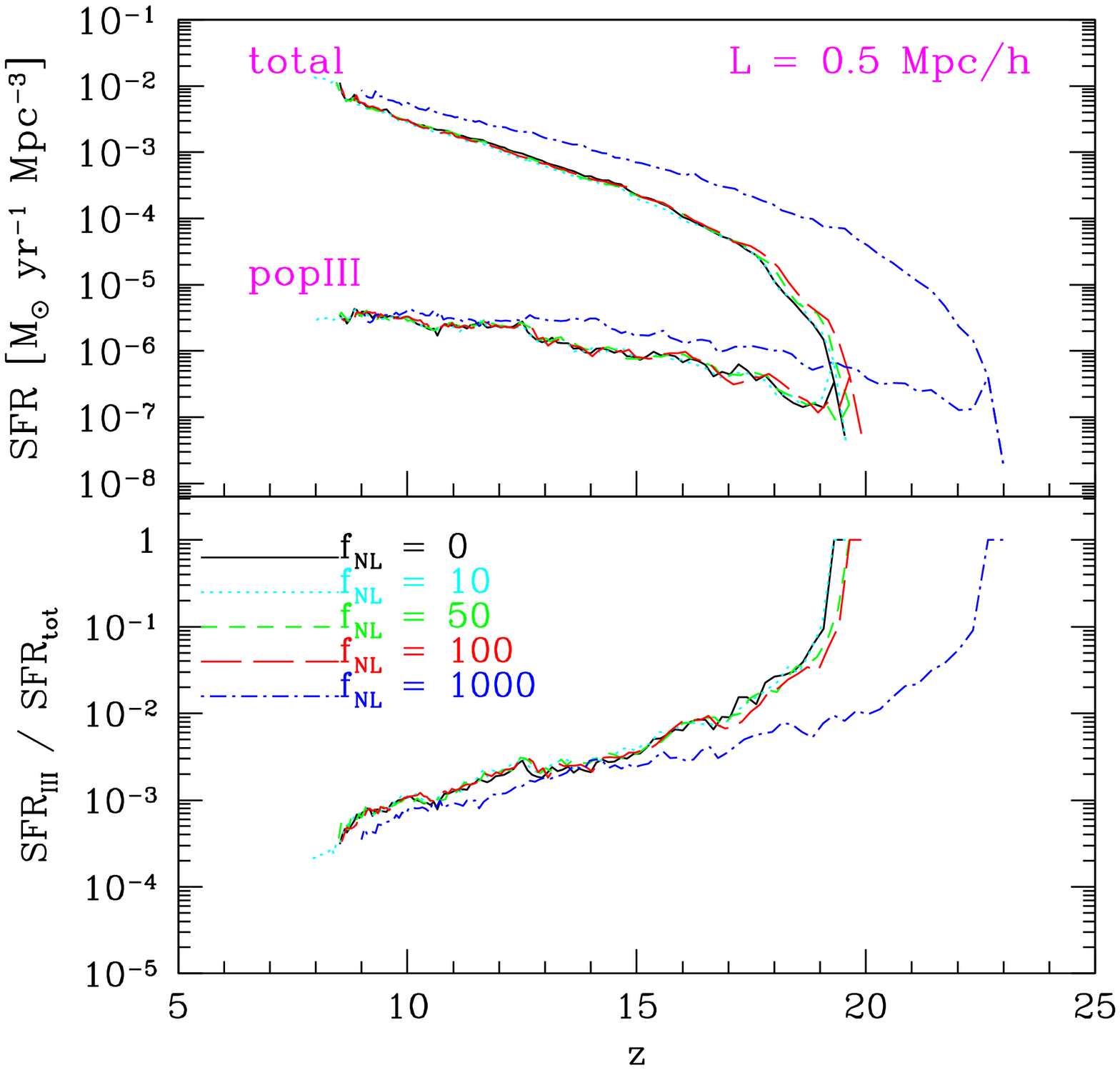}
\includegraphics[width=0.49\textwidth]{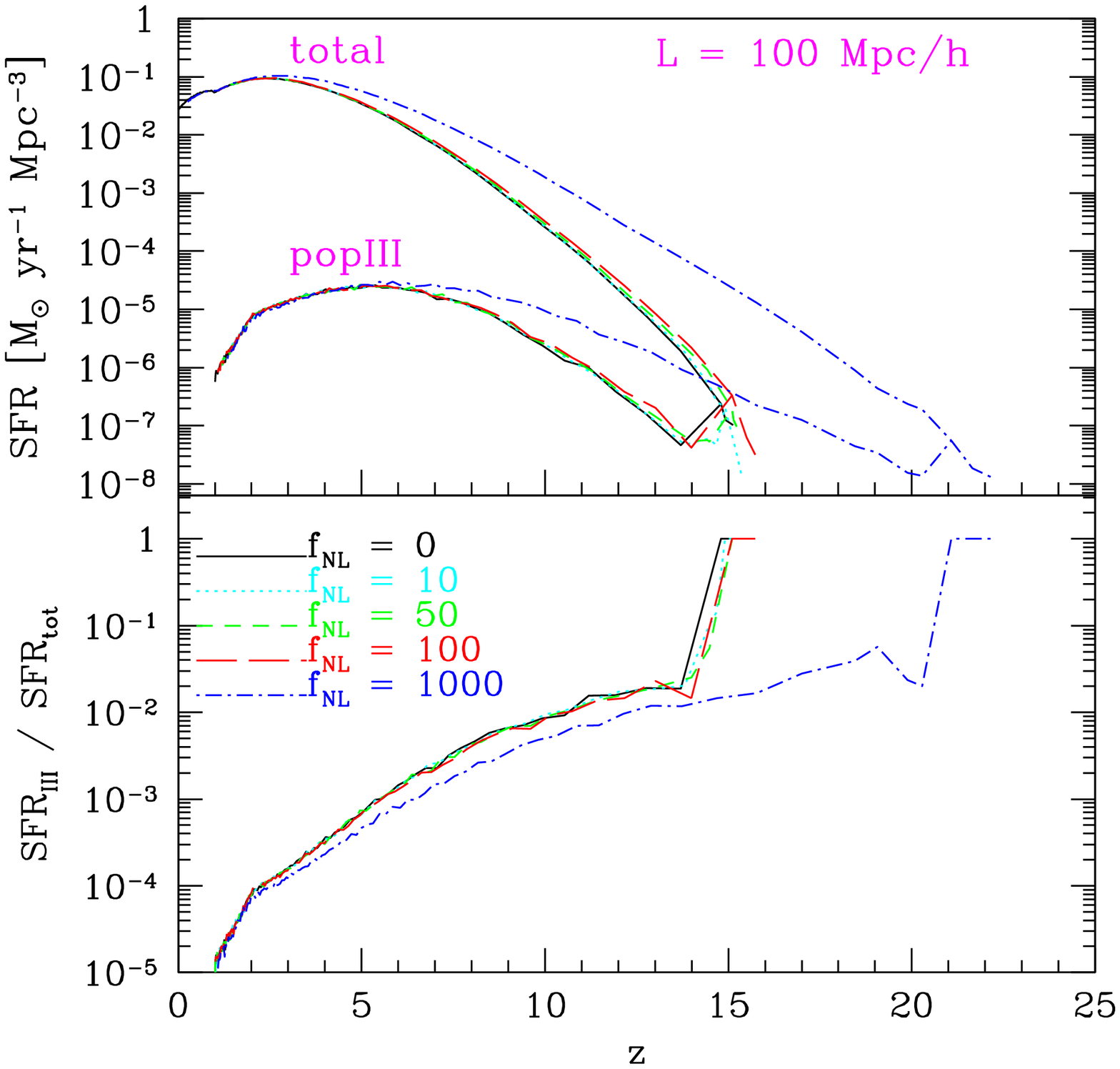}
\caption[SFR]{\small
  {\it Top panels}: 
  total and popIII star formation rate density evolution as a function of redshift, $z$, for the small (left) and the large (right) boxes. Different lines refer to \fnl~=~0 (solid black lines), 10 (dotted cyan lines), 50 (short-dashed green lines), 100 (long-dashed red lines), and 1000 (dot-dashed blue lines).
  {\it Bottom panels}: ratio between the popIII and the total star formation rate.
  The plots are taken from \citet{MaioIannuzzi2011}.
}
\label{fig:sfr}
\end{figure*}

In the present paper, we will consider a set of numerical
N-body, hydrodynamical, chemistry simulations 
with two different box sizes starting from initial
conditions with a different level of primordial non-Gaussianity.
A more detailed description of the simulations can be found in
\cite{MaioIannuzzi2011}.
Local non-Gaussianities were included in the initial conditions
by adding second-order perturbations to the
Bardeen gauge-invariant potential \cite[e.g.][]{Salopek1990}:
\begin{equation}\label{eq:nong}
\Phi = \Phi_{\rm L} + f_{\rm NL} \left[ \Phi_{\rm L}^2 - <\Phi_{\rm L}^2> \right],
\end{equation}
where $\Phi_{\rm L}$ is the {\it linear} Gaussian part, and \fnl{} the
dimensionless coupling constant controlling the magnitude of the
deviations from Gaussianity in the large-scale-structure formalism.
\\
The simulations were performed by using a modified version of the
parallel tree-PM/SPH Gadget code \cite[][]{Springel2005}, which
includes gravity and hydrodynamics, with radiative gas cooling both
from molecular and atomic (resonant and fine-structure) transitions
\cite[][]{Maio2007},
a multi-phase model for star formation \cite[][]{Springel2003},
UV background radiation \cite[][]{HaardtMadau1996},
wind feedback \cite[][]{Springel2003,Aguirre_et_al_2001}, 
chemical network for e$^-$, H, H$^+$, H$^-$, He, He$^+$, He$^{++}$, H$_2$,
H$_2^+$, D, D$^+$, HD, HeH$^+$ \cite[e.g.][ and references
therein]{Yoshida2003,Maio2006,Maio2007,Maio2009,Maio2009PhDT,Maio2010},
and metal (C, O, Mg, S, Si, Fe) pollution from popIII and/or popII-I
stellar generations, ruled by a critical metallicity threshold of
$Z_{crit}=10^{-4}\,\zsun$
\cite[see discussion in][]{Tornatore2007, Maio2010, Maio2011b}.
The cosmological parameters are fixed by assuming a flat concordance
$\Lambda$CDM model with
matter density parameter $\Omegam=0.3$,
cosmological-constant density parameter $\Omegal=0.7$,
and baryon density parameter $\Omegab=0.04$;
the present Hubble parameter is fixed to H$_0=100~h\,\rm km/s/Mpc$,
with $h=0.7$.
Finally, the matter power spectrum has a spectral index $n=1$ and 
is normalized assuming that the mass variance within 8~Mpc/{\it h}-radius 
sphere is $\sigma_8=0.9$.
\\
A Salpeter IMF with mass range [0.1, 100]~$\msun$ was adopted for the popII-I 
regime, while a top-heavy IMF 
with short-lived stars in the mass range [100, 500]~$\msun$ was assumed for 
the popIII regime (see literature for further studies on the expected range
of massive popIII stars:
\citealt{Abel2002, Yoshida2003, InayoshiOmukai2012}; 
or low-mass popIII stars:
\citealt{Yoshida2006, Yoshida_et_al_2007, CampbellLattanzio2008, SudaFujimoto2010};
and the impacts of the different assumptions: \citealt{Maio2010}).
\\
Massive stars die as SN or as pair-instability SN (PISN) in the range 
[8, 40]~$\msun$ and [140, 260]~$\msun$, respectively, polluting the 
surrounding medium and enhancing the transition from a metal-poor to a 
metal-rich regime \cite[e.g.][]{Tornatore2007, Maio2010, Maio2011b}.
Black-hole remnants form from stellar masses in the ranges 
[40, 100]~$\msun$ (popII-I progenitors), [100, 140]~$\msun$ (popIII progenitors),
and [260, 500]~$\msun$ (popIII progenitors).
\\
To follow with sufficient accuracy all the relevant scales at the different 
cosmological epochs, we consider two sets of simulations.
\\
The first one assumes small boxes with side of $\rm 0.5~Mpc/{\it h}$,
and allows us to resolve the gas behaviour down to $\sim$~pc scales
at $z \sim 9 - 30$ \cite[][]{MaioKhochfar2012}, with gas and dark-matter
mass resolutions of $42.35~\msunh$ and $275.28~\msunh$, respectively,
and comoving softening of $\rm 4~pc/{\it h}$.
\\
The box size of the second set is much larger, $\rm 100~Mpc/{\it h}$,
so that we can resolve galactic $\sim$~kpc scales at lower redshift
\cite[][]{Maio2011}, since gas and dark-matter mass resolutions are 
$3.39\times 10^8~\msunh$ and $2.20\times 10^9~\msunh$, respectively,
and the comoving softening is $7.8~\rm kpc/{\it h}$.
\\
For both sets of simulations, different levels of primordial
non-Gaussianity have been considered, namely \fnl~=~0, 10, 50, 100, and 1000.
We highlight that current data seem to suggest positive \fnl\ values, 
between $0$ and $100$ \cite[e.g.][]{Komatsu2011}, 
but in the present work we will consider the \fnl=1000 case as well, as an
extreme example.
For further details we refer to \cite{MaioIannuzzi2011}.
\\
The star formation rate for both stellar population regimes extracted from these
simulations will represent the fundamental input for our estimates of 
the GRB rates, as described in the following sections.
For the sake of clarity, in Fig.~\ref{fig:sfr} we re-propose the redshift
evolution of the star formation rate densities derived from our ten runs 
and widely discussed in \cite{MaioIannuzzi2011}.
These curves are the starting point of our following analyses.


\section{Calculation of the GRB rates}\label{Sect:results}

In the following section we will present the results for the GRB rates
expected from our simulations.
Our starting point is the comoving SFR density, $\dot\rho_\star$,
tracked by the different runs as a function of $z$ 
\cite[]{Maio2010, MaioIannuzzi2011, Maio2011cqg}, 
from which we compute the evolution of the GRB formation rate
density, $\dot n_{\rm GRB}$,
and hence the corresponding integrated GRB rate, $R$.
\\
We will proceed as follows: in the first place, we will present a
phenomenological model describing the redshift evolution of GRBs as 
observed by {\it Swift} (Sect.~\ref{sect:model});
then we will validate it against observational data at $z \ge 6$
(Sect.~\ref{sect:swiftvalidation}), i.e. the epoch when the effects 
of primordial non-Gaussianities are expected to play a major role;
and eventually we will apply it to an ideal instrument that is assumed 
to detect all the GRBs produced in the different cosmological scenarios
(Sect.~\ref{Sect:FinalResults}).

\subsection{Model description}
\label{sect:model}

The basic features of the model are presented in
Sect.~\ref{sect:formalism}, followed by the derivation of the
best-fitting values for the model free parameters in
Sect.~\ref{sect:fit}.
We stress that the parameters of the model are dependent on the whole
cosmic star formation history, and, therefore, they do depend on the
\fnl{} values, too.

\subsubsection{Formalism}
\label{sect:formalism}
The expected redshift distribution of ``observed'' GRBs can be computed 
once the GRB luminosity function (LF) and the GRB formation history have 
been specified
\cite[e.g.][]{PM2001, Firmani2004, Natarajan2005, Guetta2005, Daigne2006, Salvaterra2007, Salvaterra2009, Dai2009, WandermanPiran2010, Cao2011, Salvaterra2012CompleteSwiftSample}.
\\
We briefly recap here the adopted formalism and refer the interested reader
to \cite{Salvaterra2007, Salvaterra2009b, Salvaterra2012CompleteSwiftSample}
for more details.
\\
The observed peak photon flux, $P$, emitted by an isotropically
radiating source at redshift $z$ and corresponding luminosity distance
$d_L(z)$, as detected in the energy band $E_{\rm min} < E < E_{\rm max}$, is

\begin{equation}
P=\frac{(1+z)}{4\pi d_L^2(z)}~ \int^{(1+z)E_{\rm max}}_{(1+z)E_{\rm min}}\!\!\! S(E)~\d~E,
\end{equation}

\noindent
where $S(E)$ is the differential rest-frame photon luminosity of the
source.
To describe the typical burst spectrum we adopt a ``Band'' 
function with low- and high-energy spectral indices equal to $-1$ and
$-2.25$, respectively
\cite[see also][]{Band1993, Preece2000, Kaneko2006}.

\noindent
The spectrum normalization is obtained by imposing that the isotropic-equivalent peak luminosity is 
\begin{equation}
L=\int^{10\,\rm{MeV}}_{1\,\rm{keV}} E ~S(E) ~\d E.
\end{equation}
To estimate the peak energy of the spectrum, $E_p$, for a given $L$, 
we correlate $E_p$ and $L$ as done in
\cite{Yanetoku2004, Nava2012, Ghirlanda2012}.
\\
Given a normalized GRB LF, $\psi(L)$, the observed number rate of
bursts (in $\rm yr^{-1}$) at redshift $z$
with peak photon flux, $P$, between $P_1$ and $P_2$ is

\begin{eqnarray}
\label{eq:Ndot}
\dot N(z) \equiv \frac{\d N_{P_1<P<P_2}(z)}{\d t} & = &
\int_z^{\infty} \d z^\prime ~\frac{\d V(z^\prime)}{\d z^\prime} ~\frac{\dot n_{\rm GRB}(z^\prime)}{(1+z^\prime)} \nonumber \\
& \times & \int^{L_{P_2}(z^\prime)}_{L_{P_1}(z^\prime)} \psi(L^\prime) \d L^\prime,
\end{eqnarray}

\noindent
where the factor $(1+z^\prime)^{-1}$ accounts for cosmological time dilation,
\begin{equation}
\frac{\d V(z)}{ \d z} = \d \Omega ~ d_{c}^2(z)~\frac{c}{H(z)}
\end{equation}

\noindent
is the comoving volume element,
$\d \Omega$ is the solid angle
$d_c(z)$ is the comoving distance,
$H(z)$ is the expansion parameter
\cite[for more explicit details see e.g.][]{Weinberg1972,Hogg1999astro.ph}, 
$c$ is the speed of light,
and $\dot n_{\rm GRB}(z)$ is the actual comoving GRB formation rate 
density as a function of redshift.
\\
Here, we assume that GRBs are good tracers of star formation, and thus that
the GRB formation rate density is directly proportional to the SFR
density (see further discussion in Sect.~\ref{Sect:discussion}), i.e.
\begin{equation}
\dot n_{\rm GRB}(z) \equiv k \dot{\rho}_\star(z),
\end{equation}
where the normalization constant, $k$ (whose dimensions are the inverse of a mass), incorporates further not-well-known effects,
like GRB beaming 
\cite[][]{Frail2001, PanaitescuKumar2001, Rossi2002, Ghirlanda2007}, 
efficiencies 
\cite[][]{Fruchter2006, Yoon2006, Bissaldi2007, Soderberg2010, Grieco2012arXiv},
and black-hole production probability (depending on the adopted IMF).
\\
We will adopt (see next section for more details) a normalized GRB LF
described by a single power-law with slope $\xi$ and
decreasing exponentially below a characteristic luminosity, $L_{\star}$,
\begin{equation}\label{eq:LF}
 \psi(L) \propto \left(\frac{L}{L_{\star}}\right)^{-\xi} \exp \left(-\frac{L_{\star}}{L}\right).
\end{equation}

\noindent
Then, we consider the possibility that the GRB LF evolves by setting
$L_{\star}(z)=L_{{\star},0}(1+z)^\delta$, where 
$L_{{\star},0}$ is the characteristic luminosity at $z=0$,
and $\delta$ is the evolution parameter.
\\
For simplicity, the normalization of $\psi(L)$ is included in $k$, and it is fixed
when the GRB number rate in equation~(\ref{eq:Ndot}) is normalized to the rate
observed at $z=0$.
\\
From the previous relations we can finally compute the GRB rate
(in units of yr$^{-1}$~sr$^{-1}$), $R$, as:

\begin{equation}
\label{eq:R}
R(z) = \frac{\d \dot N(z)}{\d \Omega},
\end{equation}
i.e. by taking the derivative with respect to the solid angle of the
GRB number rate in equation~(\ref{eq:Ndot}).

\begin{table*}
  \begin{center}
    \begin{tabular}{lcccccccr}
      \hline
      \hline
      Model & $\log_{10}(~k~[\msun^{-1}]~)$ &  $L_{\star,0,51}$ & $\xi$ & $\delta$ & C-stat & \\
      \hline
      f$_{\rm NL}$=0 & $-7.70_{-0.06}^{+0.09}$ &  $0.17_{-0.10}^{+0.23}$ &
      $2.04^{+0.15}_{-0.11}$ & $2.59_{-0.57}^{+0.63}$ & 29  \\
      f$_{\rm NL}$=10 & $-7.70_{-0.06}^{+0.09}$ & $0.18_{-0.10}^{+0.24}$ &
      $2.04^{+0.15}_{-0.11}$ & $2.57_{-0.57}^{+0.62}$ & 29 \\
      f$_{\rm NL}$=50  &  $-7.70_{-0.06}^{+0.09}$ & $0.18_{-0.10}^{+0.24}$ &
      $2.04^{+0.15}_{-0.11}$ & $2.57_{-0.57}^{+0.62}$ & 29 \\
      f$_{\rm NL}$=100  &  $-7.70_{-0.06}^{+0.09}$ & $0.19_{-0.11}^{+0.25}$ &
      $2.04^{+0.15}_{-0.11}$ & $2.53_{-0.56}^{+0.62} $ & 30 \\
      f$_{\rm NL}$=1000 & $-7.75_{-0.06}^{+0.10}$ & $0.36_{-0.21}^{+0.48}$ &
      $2.10_{-0.13}^{+0.20}$ & $2.08_{-0.50}^{+0.56}$ & 30 \\
      \hline
    \end{tabular}
    \caption{
      Best-fit values and 1-$\sigma$ errors for the free parameters
      of the GRB model, computed for the different cosmologies.
      From left to right the columns refer to the value of:
      \fnl;
      the GRB normalization in [$\msun^{-1}$], $k$;
      the characteristic luminosity at $z=0$ in [$10^{51}\,\rm~erg~s^{-1}$], $L_{\star,0,51}$;
      the slope parameter of the GRB LF, $\xi$;
      the redshift evolution parameter of the characteristic luminosity in the GRB LF, $\delta$;
      the total C-stat value (i.e., the sum of the C-stat values obtained
      from the fit of the BATSE and {\it Swift} dataset) -- for more details see
      \citet{Salvaterra2012CompleteSwiftSample}.
      The total number of data points used to perform the fit is 33.
    }
  \end{center}
\end{table*}

\subsubsection{Parameter estimation} 
\label{sect:fit}

\begin{figure}
  \centering
  \includegraphics[width=0.45\textwidth]{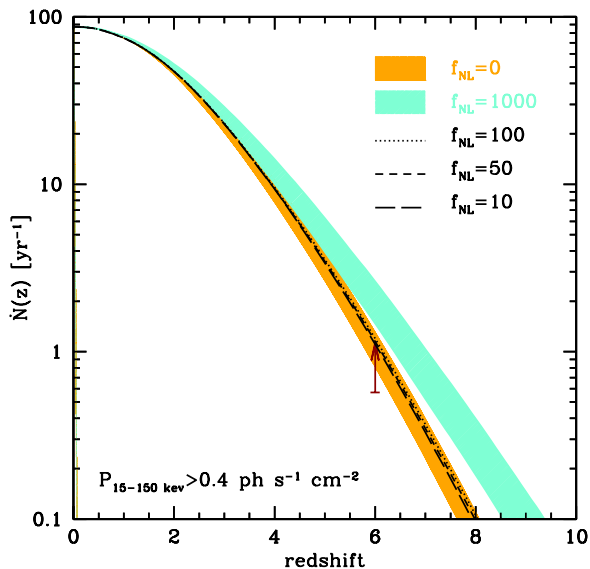}
  \caption[Swift sensitivity]{\small 
    Redshift evolution of the observed
    number rate of GRBs, $\dot N$, at the sensitivity of {\it Swift} instrument,
    corresponding to a peak flux of 0.4 ph s$^{-1}$ cm$^{-2}$ in
    the [15, 150]~keV band, and for the {\it Swift} field of view
    $\Delta\Omega_s = 1.4\,\rm sr$. Model results for f$_{\rm NL}=0$
    and f$_{\rm NL}$=1000 are shown as dark orange and light blue
    shaded regions, respectively, taking into account the errors on the
    evolution parameter.
    The trends (without errors) for the models with \fnl~= 10, 50, 100 are
    shown with long-dashed, short-dashed, and dotted lines, respectively.
    The arrow refers to the lower limit on the rate of
    GRBs at $z>6$, 
    imposed by the four confirmed detections at these redshifts (see text).
  }
  \label{fig:swift}
\end{figure}

The values of the free parameters of the model 
(i.e. $L_{\star,0}$, $\xi$, $k$ and $\delta$) 
are optimized separately for all the models,
by using the SFRs obtained from the different cases 
in the 100~\Mpch-size boxes.
\\
We proceed by minimizing the C-stat function \cite[][]{Cash1979},
jointly fitting the observed differential number counts in the
$[50, 300]~\rm keV$ band of BATSE \cite[][]{Stern2001} and the
observed redshift distribution of bursts in a redshift complete
subsample between $z=0.13$ and $z=5.47$ of {\it Swift} bursts 
with photon fluxes in excess of 2.6~ph~s$^{-1}$~cm$^{-2}$ in the 
{\it Swift} $[15, 150]~\rm keV$ band
\cite[for more details see][]{Salvaterra2012CompleteSwiftSample}.
\\
While the redshift complete {\it Swift} subsample provides a powerful
test for the existence and the redshift evolution of the long
GRB population, the fit to the BATSE number counts allows to obtain
the normalization $k$ and to better constrain the GRB LF free
parameters.  It is worth noting that the same best-fit parameters
provide a good fit also to the {\it Swift} differential peak-flux
number counts once the energy band ($[15, 150]~\rm keV$), the field
of view ($\Delta \Omega_s = 1.4~\rm sr $), and the observing lifetime
of {\it Swift} are considered \cite[][]{Salvaterra2007}.
\\
The best-fit values together with their 1-$\sigma$ confidence
levels are provided for the different values of \fnl{} in
Table~\ref{tab:parameters}.
\\
We note that since the star formation rate densities are similar for
\fnl~$\le 100$, the best-fit parameters obtained do not differ
significantly with respect to the Gaussian case.
Also in the most extreme case \fnl=1000 the GRB LF best-fit parameters are still
consistent with those obtained in the Gaussian cosmology.
However, in this case, the normalization $k$ and the evolution parameter $\delta$
are affected by the different shape of the cosmic SFR.
This was indeed expected: because of the enhanced SFR at high redshift
in the \fnl=1000 model, a slightly lower evolution is required to reproduce the
observed redshift distribution of the complete sample of bright {\it
Swift} GRBs and, consequently, also a different normalization is found.
\\
The LF of popIII GRBs is completely unknown. 
To compute their rate, we follow \cite{Campisi2011} and assume
that popIII GRBs can be described by equation~(\ref{eq:LF})
with $\xi=1.7$ and $L_{\star}=10^{54}~\rm erg s^{-1}$ constant in
redshift (i.e. $L_{\star, 0}=10^{54}~\rm erg s^{-1}$ and $\delta = 0$)
\cite[][]{Toma2011}.
The normalization is then obtained by imposing that none of
the $\sim 500$ GRBs detected by {\it Swift} so far were powered
by popIII star explosions. We checked that our results do not change
significantly when varying $\xi$ between 1.5 and 2 and 
$\log (L_{\star, 0}/\rm erg~s^{-1})$ between 53 and 55.


\subsection{Validation from the  {\it Swift} redshift distribution}
\label{sect:swiftvalidation}

Before calculating the GRB rate expected for different cosmologies,
we test the validity of our theoretical model by means of the 
{\it Swift} data.
We remind that, as of today, the {\it Swift} instrument has detected 
604 GRBs in a lifespan of about 7~years, and the redshift complete 
(sub-)samples that have been extracted so far have only several tens
of data points \cite[see][]{Perley2009,Greiner2011, Salvaterra2012CompleteSwiftSample, Hjorth2012}.
\\
Fig.~\ref{fig:swift} reports the redshift evolution of all models expected
at the {\it Swift} sensitivity, corresponding to a peak photon flux
of 0.4~ph~s$^{-1}$~cm$^{-2}$ in the $[15, 150]~\rm~keV$ band.
The {\it Swift} field of view of $\Delta \Omega_s = 1.4~\rm sr$ 
has been assumed.
If we compare the \fnl=0 and 1000 cases, it is evident that significant 
differences arise at $z\gsim 6$, where $\dot{N}$ changes by at least
factor of $\sim 2$.
At lower redshift the two distributions are very similar and possible 
differences fall within the uncertainties (shaded regions) on the 
evolution parameter, $\delta$.
Indeed, the upper and lower bounds of the shaded regions correspond to 
the evolution obtained by fitting the complete {\it Swift} sample with 
the maximum and minimum values of $\delta$ as quoted in 
Table~\ref{tab:parameters}.
We note that in principle an instrument like {\it Swift} can distinguish 
between a Gaussian and a highly non-Gaussian model simply on the basis 
of the rate of GRB detections at high $z$.
\\
The four confirmed detections at 
$z>6$ (GRB~050904 at $z=6.3$ by \citealt{Kawai2006}; 
GRB~080913 at $z=6.7$ by \citealt{Greiner2009};
GRB~090423 at $z=8.2$ by \citealt{Salvaterra2009},\citealt{Tanvir2009}; 
GRB~090429B at $z\simeq 9.4$ by \citealt{Cucchiara2011})
correspond to a rate of
$\dot N(6) = 0.57\pm 0.28$ GRBs per year,
derived by using the entire timespan of {\it Swift} ($\sim 7$~years).
At face value, this is fully consistent with the predictions that our 
model provides for the Gaussian case.
Moreover, since the GRB redshift distribution for mildly non-Gaussian 
models does not differ significantly in the redshift range probed by
{\it Swift}, the observed high-$z$ rate is also consistent with any
non-Gaussian model with a positive but smaller than $\sim 100$ \fnl.
\\
However, we have to remind that the observed value for $\dot N(z)$ at
$z=6$ is a lower limit for the rate of high-$z$
GRB detections with {\it Swift}, since some bursts at $z>6$
could be hidden among the large fraction ($\sim 2/3$) of 
GRBs for which the redshift has not been measured.
For this reason, the previous constraint seems to rule out very negative 
values of \fnl{} that would fall below the aforementioned limit.
\\
A strong upper limit of $\le 14\%$ on the fraction of $z>6$ GRBs detected 
by {\it Swift} has been recently determined by  \cite{Jakobsson2012}\footnote{
We stress that the upper limit at $z=6$ of $14\%$ suggested by 
\cite{Jakobsson2012} refers to a subsample of 69 {\it Swift} bursts
and is obtained by assuming that the GRBs that could not be identified as 
low-redshift are actually at $z \ge 6$.
Thus, the value of $14\%$ must be taken as a very strong upper limit.
}.
Considering the 604 GRBs constituting the current {\it Swift} sample, this 
corresponds to at most 85 GRBs at $z>6$, and to a rate of
$\dot N(6) \lesssim (12 \pm 1)~\rm yr^{-1}$
(as this value is quite large, we do not show it in Fig.~\ref{fig:swift}).
\\
The \fnl=1000 case and larger values, already excluded 
by CMB analyses \cite[][]{Komatsu2011},
lie off the observed rate by one order of magnitude or more.
\\
The previous considerations are based on the data point for the GRB rate 
at $z\simeq 6$, but higher-$z$ data, and larger, redshift complete samples
\cite[e.g.][]{Perley2009,Greiner2011,Salvaterra2012CompleteSwiftSample,
Hjorth2012} together with a better knowledge of the GRB luminosity
function and of its redshift evolution, are needed in order to reduce
error bars and to put tighter constraints on the amount of primordial
non-Gaussianity on the basis of the observed GRB redshift distribution.
More precisely, in order to discriminate, at redshift $z = 6$,
between e.g. the \fnl$=0$ and the \fnl$=1000$ cases
(whose number rates at $z=6$ differ by a factor of $\approx 2$),
with a $3\sigma$ confidence level, one should have a redshift complete 
sample of roughly 800 GRBs (Poissonian errors have been assumed,
and a current {\it Swift} number rate of $\dot N(0) = 88~\rm yr^{-1}$,
according to Fig.~\ref{fig:swift}).
Such a large sample would also allow us to constrain the
GRB LF quite accurately and then strongly reduce the error bars.
A confidence level of $1\sigma$ would require a smaller redshift 
complete sample of about 200 GRBs.


\subsection{Resulting GRB rates from the simulated cosmologies}
\label{Sect:FinalResults}

After the validation of the model (Sect.~\ref{sect:swiftvalidation})
performed by exploiting the calibration of the GRB rates against
{\it Swift} data (Sect.~\ref{sect:formalism} and \ref{sect:fit}),
we now apply it to different non-Gaussian cosmologies to derive
count predictions as a function of $z$.
Note that we will assume the normalization derived from the
simulations with 100~\Mpch-side boxes for the small
0.5~\Mpch-side boxes, as well.  In fact, the latter are not run down to $z \sim 0$
and thus can not be used for normalization purposes.
\\
We stress that the following results are obtained by assuming an
ideal instrument, that is able to detect all the GRBs
at high redshift.
This is important to note, because, independently
of the overall normalization, the main effects of primordial non-Gaussianities
on GRBs are expected to be originated by the differences shown in the redshift
evolution of the SFRs \cite[see][]{MaioIannuzzi2011} and, hence, in
the different GRB rates in the various models.

\begin{figure*}
\centering
\includegraphics[width=0.49\textwidth]{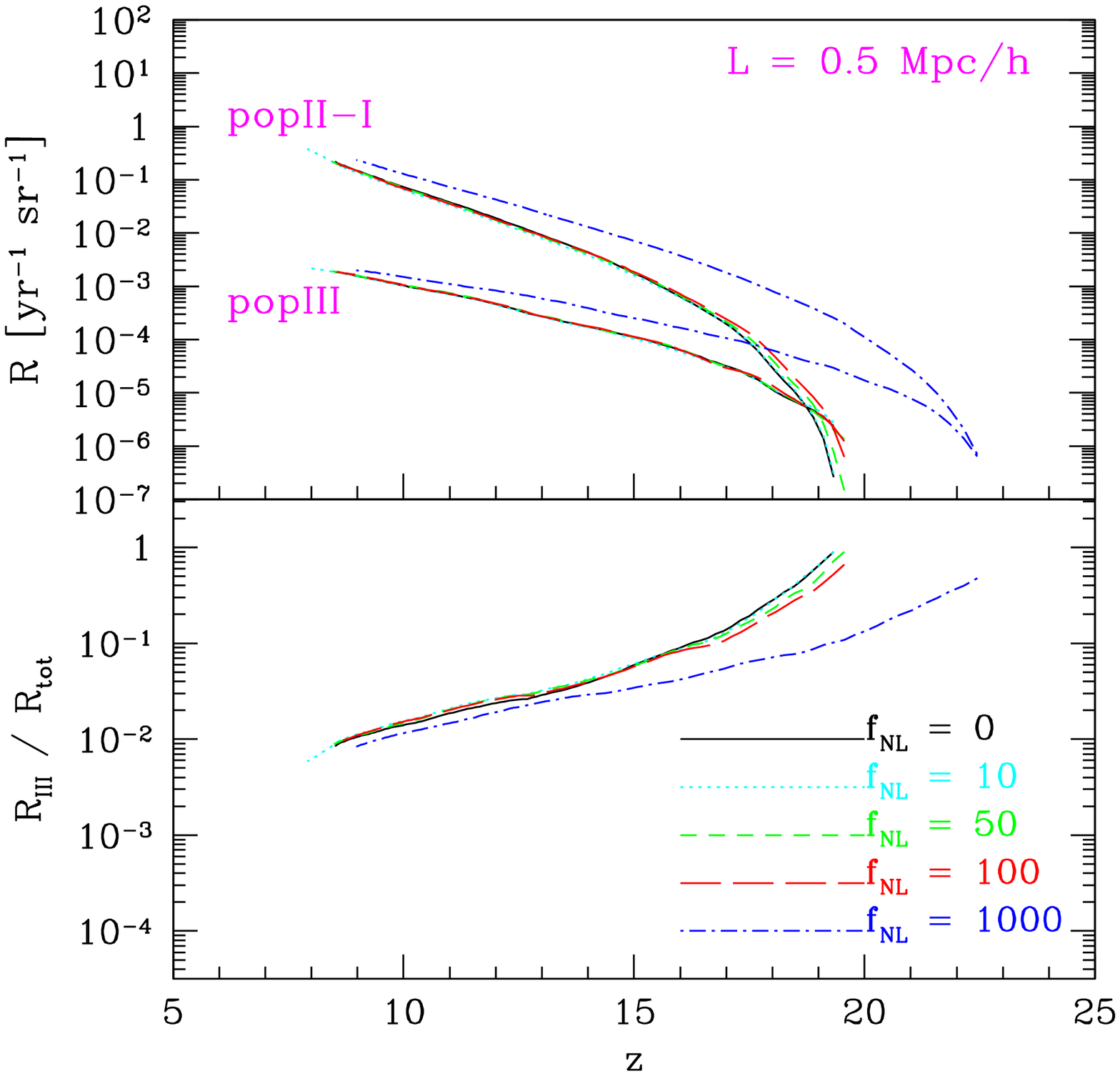}
\includegraphics[width=0.49\textwidth]{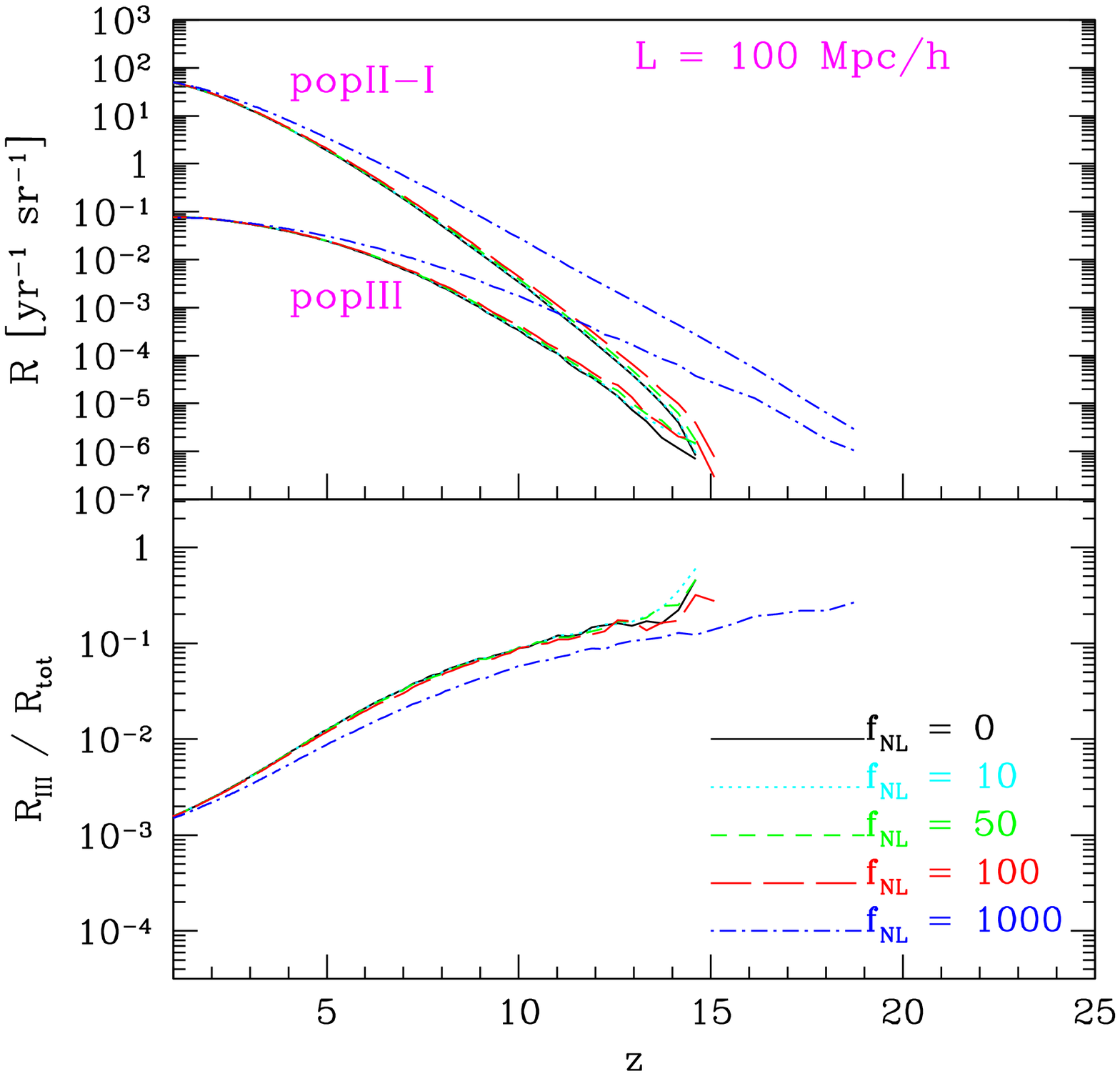}
\caption[Redshift comparison]{\small 
  {\it Top panels}: The expected
  popII-I and popIII GRB rates, $R$, in the $0.5\,\Mpch$-side boxes
  (left), and in the $100\,\Mpch$-side boxes (right), for models with
  different primordial non-Gaussianity: \fnl = 0 (solid black line),
  10 (dotted cyan line), 50 (short-dashed green line), 100
  (long-dashed red line), and 1000 (dot-dashed blue line).  {\it Bottom
    panels}: The corresponding relative contributions of the popIII
  GRB rates, $R_{\rm III}$, to the total rate, $R$, for the different
  cosmological models.
}
\label{fig:Rall}
\end{figure*}

\subsubsection{Evolution of the GRB rates}

In Fig.~\ref{fig:Rall} we plot the GRB rate, $R$, for the small
0.5~\Mpch-size boxes (left panels) and the large 100~\Mpch-size boxes
(right panels).
In the top panels, we show the redshift evolution for
all the \fnl{} scenarios considered, while in the bottom panels we focus
on the relative contribution of the popIII GRB rate ($R_{\rm III}$) to
the total rate, that is widely dominated by popII-I stellar
generations, at $z\lesssim 20$.
\\
Besides small differences for the onset times of star formation, due
to the different resolutions of the 0.5 and 100~\Mpch-side boxes
\cite[see details on resolution issues in][]{Maio2010,
  MaioIannuzzi2011}, in both small and large volumes the effects due
to the presence of primordial non-Gaussianities are visible at $z
\gtrsim 10-15$,
while the rates eventually converge at later times, when feedback
mechanisms start dominating the gas behaviour and the resulting star
formation.
\\
In the small boxes, deviations from the Gaussian case are evident at
earlier times, because these simulations can sample the very small
primordial mini-haloes, which are extremely sensitive to the
underlying matter distribution (top-left panel).
As a consequence, star formation is resolved already at very high redshift 
and GRB rates of the order of $\sim 10^{-6}\,\rm yr^{-1}~sr^{-1}$ are
expected at redshifts as high as $z\sim 23$ for \fnl$=1000$, and
$z\sim 19-20$ for \fnl$=0-100$. Similar values are reached in the
large 100~\Mpch-side boxes only at 
$z\sim 20$ for \fnl$=1000$, and $z\sim 15$ for \fnl$=0-100$.
\\
These trends are valid for both popII-I and popIII regimes, even
though the latter is usually negligible, predicting popIII GRB
rates, $R_{\rm III}$, that, following the behaviour of the popIII SFRs, drop
by two orders of magnitude (bottom-left panel).
\\
The larger boxes miss the very small primordial haloes because of
lack of resolution, but can sample much larger scales, showing that the
effects of primordial non-Gaussianity can still be present at redshift
$z\sim 5-10$ (top-right panel), i.e. for the whole first billion years
of the Universe, when the GRB rates should be only one or two orders
of magnitude smaller than at present time.  Also in these boxes,
significant differences in the GRB rates are found only between the
\fnl~$=0$ and \fnl~$=1000$ scenarios.  This holds for the
corresponding popIII contributions (bottom-right panel), as well, and
is consistent with what found in the smaller boxes, and with the
converging behaviours at redshift below $\sim 6$.

\begin{figure*}
  \centering
  \includegraphics[width=0.49\textwidth]{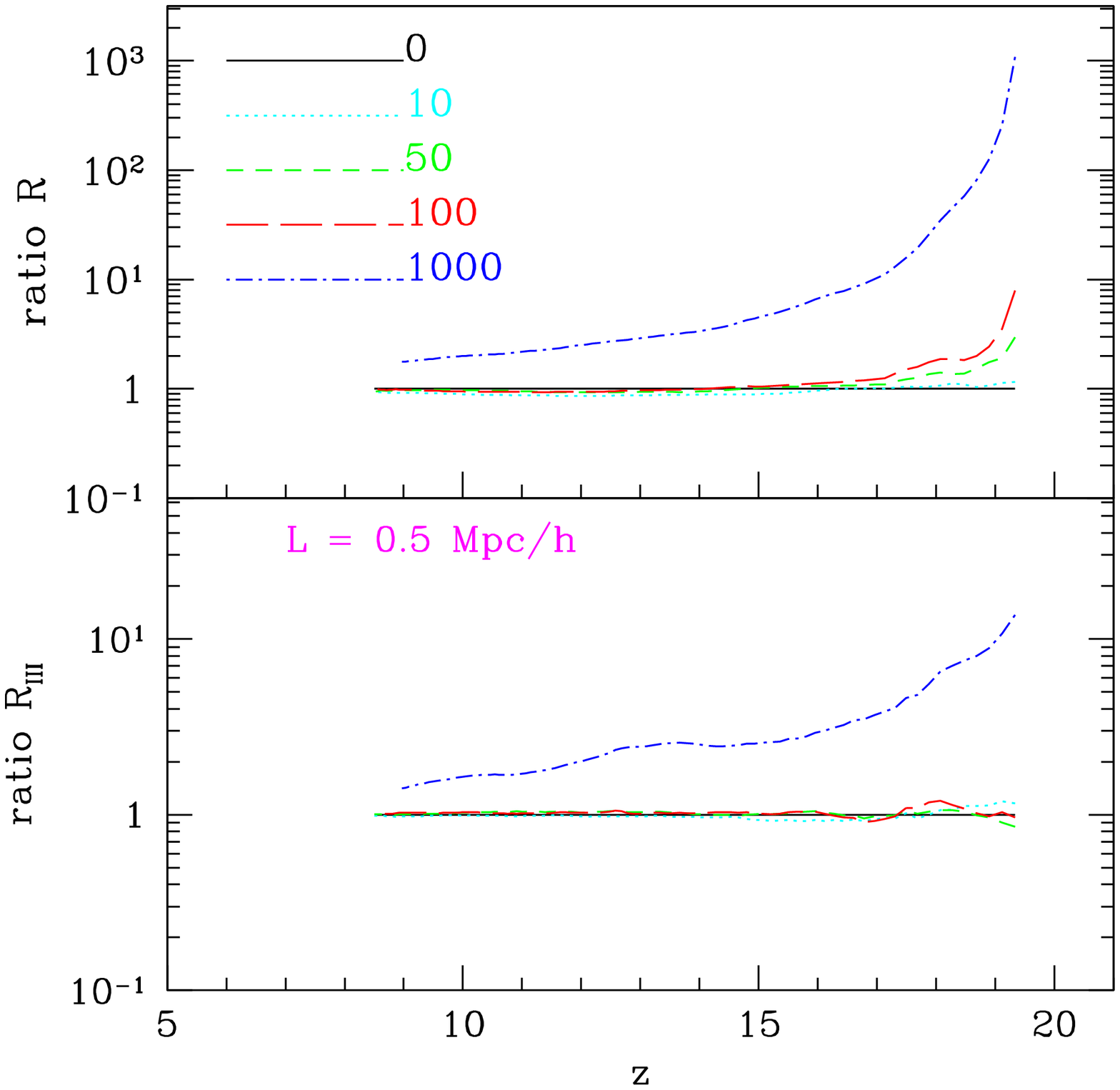}
  \includegraphics[width=0.49\textwidth]{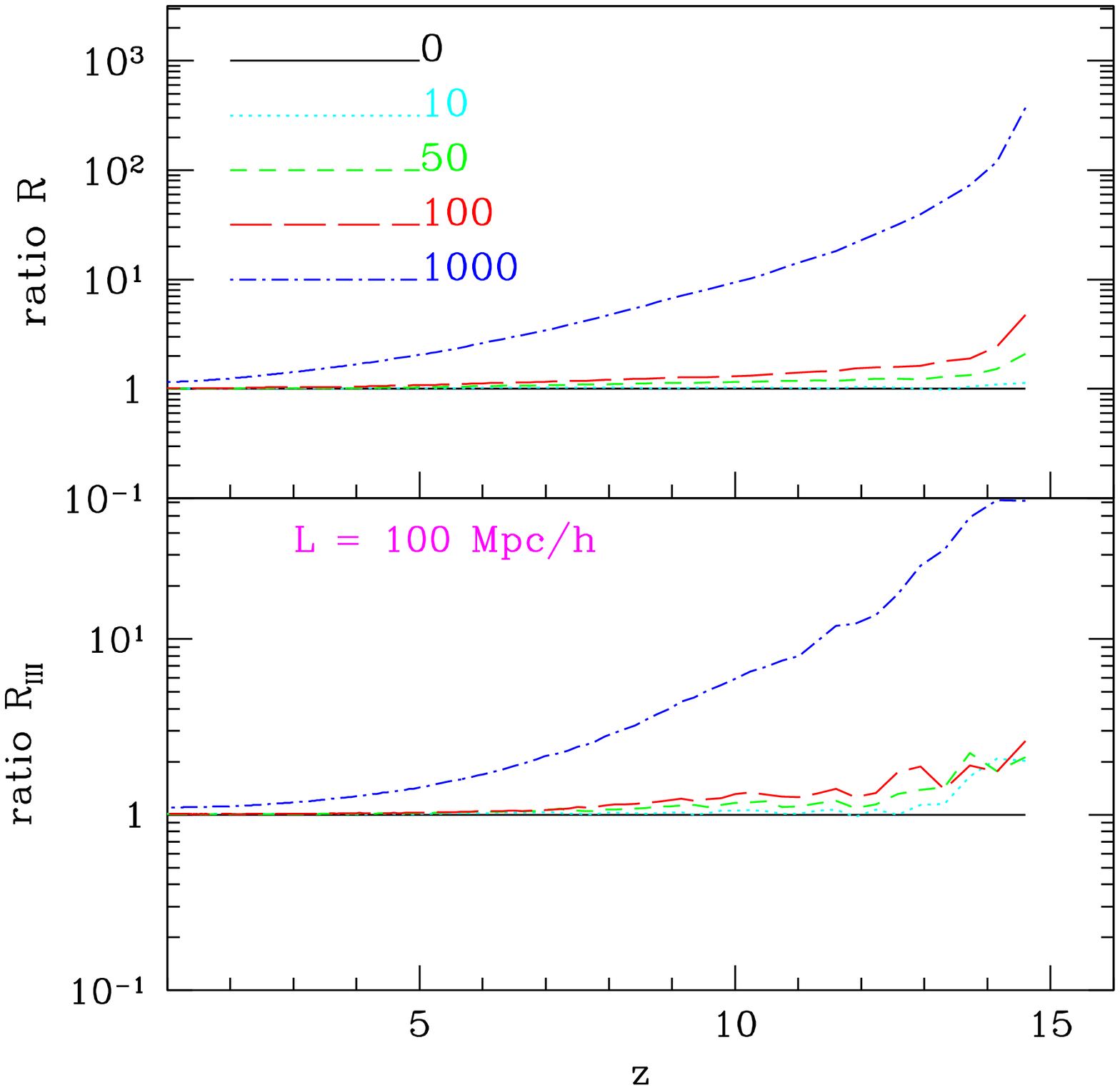}
  \caption[Redshift comparison]{\small 
    {\it Top panels}: The ratio between the popII-I GRB rates predicted 
    for the different non-Gaussian models and the Gaussian case.
    Results for the $0.5\,\Mpch$-side boxes and the $100\,\Mpch$-side 
    boxes are shown in the left and right panels, respectively.
    Different lines refer to \fnl = 0 (solid black line), 10 (dotted cyan line),
    50 (short-dashed green line), 100 (long-dashed red line), 1000
    (dot-dashed blue line).
    {\it Bottom panels}: The corresponding ratios for the popIII GRB rates.}
  \label{fig:Rcomparefnl}
\end{figure*}

\subsubsection{Comparison of the Gaussian and non-Gaussian models}

\noindent
To directly compare and isolate non-Gaussian effects, in the upper
panels of Fig.~\ref{fig:Rcomparefnl}, we plot the ratios between the
results for the different non-Gaussian cases and the
Gaussian model (\fnl=0), for both the 0.5~\Mpch (left panels) and
100~\Mpch (right panels) side boxes.
Effects of large non-Gaussianities (\fnl~$=1000$) are very well visible 
at almost any redshift with a rate that is boosted by about $\sim 3$ orders 
of magnitude in all boxes, at early epochs.  This is due to the
fact that in such models, over-densities are heavily biased to larger
values, and, therefore, induce an earlier onset of star formation.
More precisely, small scales (left panels) seem to depend very tightly
on the underlying matter distribution, with enhancements of the GRB
rate at $z\sim 20$ by a factor of $\sim 10^3$, $ 10$, $ 3$, and a few
per cent, for \fnl~$=~$1000, 100, 50, and 10 respectively.  In the
latter three cases, feedback mechanisms from on-going star formation
are able to reshuffle the gas and drive its hydrodynamical
behaviour.
As a consequence, the non-Gaussian effects below Mpc scales
are almost washed out by redshift $z\sim 15$.
The highly non-Gaussian case (\fnl=1000), instead, shows more prolonged 
effects, with variations by a factor $\sim 10$ at $z\sim 15$, and a factor 
of $\sim 2$ at $z\lesssim 10$.
\\
On larger scales (right panels), the ratios are similar for the
various non-Gaussian cases, with corresponding delays toward lower
redshift in the low-\fnl{} scenarios.
\\
As a conclusion, we can state that the presence of primordial
non-Gaussianities in the density fluctuations enhances early GRB rates
and has effects up to $z\sim 10$ on Mpc scales, and at least $z\sim 5$ 
on much larger scales.
\\
To check whether different stellar populations can have
different contributions, we can compare the corresponding ratios for
popIII GRB rates only.
The bottom panels in Fig.~\ref{fig:Rcomparefnl} readily demonstrate
that the popIII GRB rates are less indicative of primordial
non-Gaussianities, and less powerful in discriminating different
\fnl{} scenarios, mostly for \fnl~$\lesssim 100$.
The fundamental
reason is that the popIII contribution to the SFRs is very
noisy due to the short life-times involved \cite[][]{Maio2010}, and 
thus also the corresponding contribution to the GRB rates present more 
irregularities compared to the total (popII-I) GRB rates.
\\
Finally, we checked that uncertainties in the unknown popIII IMF\footnote{
  Here we adopted alternatively, as an extreme case, a Salpeter-like
  popIII IMF.
},
in the $Z_{crit}$ value, and in the stellar yields would not lead to
significant differences for the previous results
\cite[see also related discussions in][]{Maio2010, MaioIannuzzi2011}.
Similarly, changes in the popIII GRB efficiency do not alter the relative 
effects of non-Gaussianities, since they would correspond just to a different
overall normalization.


\section{Discussion and conclusions}\label{Sect:discussion}
In this work we have discussed the possibility of using GRBs as
possible probe of the presence of primordial non-Gaussianities in the
density field. This has been done using the outputs of two sets of 
N-body, hydrodynamic, chemistry simulations presented in 
\cite{MaioIannuzzi2011} (as also briefly described in 
Sect.~\ref{Sect:simulations}).
\\
Besides gravity and hydrodynamics, the runs include radiative gas cooling 
both from molecules and atomic (resonant and fine-structure) transitions, 
star formation, UV background, wind feedback, and chemistry evolution
for various metal species, for both population~III and 
population~II-I stellar generations.
\\
Assuming that long $\gamma$-ray bursts are fair tracers of star
formation \cite[as suggested by e.g.][]{Nuza2007, Lapi2008, Levesque2010, 
Campisi2011b, Mannucci2011, Sanders2012arXiv, Michalowski2012arXiv},
we propose to use them as probes of the underlying matter distribution at 
high redshift, when the possible presence of non-Gaussianity would have 
the strongest visible effects on the baryon evolution.
\\
By validating our calculations of the GRB rate against {\it Swift} data, 
we are able to exclude from the non-linearity parameter space very 
negative values for \fnl{}
\cite[consistently with independent results from CMB data, e.g.][]{Komatsu2011}.
\\
When applying our model to different non-Gaussian scenarios,
we find that already at $z\gtrsim 6$ cosmologies with large \fnl{}
values present distinctive characteristics compared to those of the Gaussian
case, independently from the errors on the LF parameter estimates.
Both on large and small scales, at very early times ($z\sim 15-20$) 
the boost in the rate due to non-Gaussianities is $\sim 2 - 3$ orders
of magnitudes for \fnl=1000, and up to a factor of $\sim 10$ for \fnl=100.
Differences of a factor of $\sim 2$ are still visible for milder values
(\fnl $\sim$ 50).
However, while at small scales we find quick converging trends at lower redshift 
($z\sim 9$), determined by the locally on-going star formation and feedback 
episodes, larger-scale volumes sample bigger objects and thus can retain memory 
of the primordial matter distribution even at $z\sim 5-10$.
\\
These effects are particularly evident on the total GRB rate, that is
largely dominated by popII-I stars, while the result for the popIII
GRB rate is noisier, mostly for \fnl$\sim 0-100$, as a consequence
of the corresponding, short-lived, popIII star forming regime
\cite[][]{Maio2010}.
\\
Additional changes in the popIII IMF, yields, $\Zcrit$, or the
overall normalization of the GRB rates will not alter these findings
\cite[see also more discussion in][]{Maio2010}.
\\
We have to recall that, when estimating the level of primordial
non-Gaussianity, some difficulties come from the well known
degeneracies of \fnl{} with other factors, like cosmological parameters
(e.g. the power spectrum normalization $\sigma_8$, or the
equation-of-state parameter $w$), or with higher-order effects coming
from baryonic matter evolution \cite[e.g. supersonic bulk flows at
early times;][]{Tseliakhovich2010, Maio2011}.
\\
We also warn the reader that the main assumption underlying our formalism
is that GRBs are {\it unbiased} tracers of star formation
\cite[][]{Fynbo2003, Fynbo2009, Stanek2006, Modjaz2008, LevesqueI2010, LevesqueII2010}.
Despite this has been recently supported by several works (see above),
arguments for the existence of some possible biases exist in the
literature, in particular linked to metallicity selection of the host
galaxies \citep[e.g.][]{Langer2006}.
Such effects could alter the intrinsic redshift distribution of GRBs
\cite[e.g.][]{Natarajan2005, Langer2006, Salvaterra2007, Cao2011, Salvaterra2012CompleteSwiftSample}.
However, we do not expect this to have significant impacts on the
estimated trends of primordial non-Gaussianities.
Indeed, any metallicity bias for the 
GRB formation is not supposed to be too strong, i.e. possible metallicity 
thresholds for the GRB progenitor stars can not be much lower 
than $\sim 0.3\;Z_\odot $ \cite[see][]{Campisi2011}.
\\
As shown in this paper, the differences in the GRB rate induced
by non-Gaussianities are expected to be significant at very high
redshift.
At $z > 6$ most of the galaxies \cite[][]{Salvaterra2011} and, in particular, 
most of the GRB progenitors \cite[][]{Salvaterra2012} have metallicities 
below this threshold \cite[see also detailed studies in][]{Maio2010}.
These studies find that only a small fraction ($\lesssim 5\%$)
of galaxies at $z=6$ has got a metallicity $Z \ge 0.3~\zsun$,
rapidly decreasing at higher redshift.
Therefore, at least at these early times, our 
assumption of GRBs as fair tracers of the cosmic SFR is quite solid. 
Furthermore, we checked that the difference among Gaussian and 
non-Gausssian models remains unchanged when galaxies with metallicities 
larger than $0.3~\zsun$ were excluded from our analyses.
\\
We note that estimates of non-Gaussianities \cite[e.g.][]{Komatsu2011}
based on cosmic microwave background and large-scale-structure data
seem to support positive \fnl{} values up to ~$\sim 100$. 
This implies that at early epochs we expect an enhancement of the GRB rate 
up to a factor of 10 with respect to the standard Gaussian case.
\\
We stress that the existence of GRBs at such high redshift is not
unlikely, as they are tightly linked to star formation episodes.
In principle, they could be observable thanks to their large intrinsic
luminosity and longer time dilution of the afterglow.
None the less, from an observational point of view, detections of GRB 
afterglows at very high redshift are complicated by Lyman-$\alpha$ 
absorption from inter-galactic gas.
In fact, for bursts at $z\ge 15$, as the ones we
are interested in here, no flux can be detected in photometric bands
bluer than the K~band (at $\sim 2.2~\mu m$). At $z>18$, where the
largest differences between Gaussian and mildly non-Gaussian models are
expected, observations in the infra-red band are needed. Since
follow-up observations of GRB afterglow are generally carried out in
optical-NIR bands, extreme high-$z$ GRBs can be missed.
However, a small population of extremely dark GRBs 
\cite[e.g.][]{Greiner2011}, i.e. bursts for which the
afterglow remains undetected in spite of early and deep K~band
observations, has been recently identified \cite[][]{DEliaStratta2011}.
While the nature of these GRBs is still matter of debate
and alternative explanations for their darkness do exist\footnote{
Different possible explanations have been proposed, ranging from 
extreme dust absorptions at $z=0$ to less extreme, but still demanding, 
extinctions at $z\simeq 4-5$.
Other suggestions invoke a complete revision of the dust extinction curves 
or even more exotic non-standard models for the GRB afterglow emission
\cite[for a more detailed discussion see][]{DEliaStratta2011}.
},
it is possible that these bursts (or at least one of them) are at $z\ge 18$.
If confirmed, this could provide evidence in favor of a mildly positive
non-Gaussian parameter (\fnl{} in the range $10-100$, see Fig.~\ref{fig:Rall}).
Future detections of extremely dark GRBs 
\cite[as the ones by][]{DEliaStratta2011}
at redshift $z\gtrsim 20$ and with a substantial rate,
of at least $\sim 10^{-6}\,\rm yr^{-1}~sr^{-1}$, might be an indication of
even bigger values for \fnl.
Naively speaking, a determination of the rate for such GRB would lead to 
about $\sim (0.1\pm 0.1)~\rm yr^{-1} sr^{-1}$, but one should also consider 
that the probability of observing such event is almost as small as
$\sim 10^{-3}$, since this is a unique case out of the 604 {\it Swift} GRBs.
In principle, this would imply positive \fnl{} values, but with huge error
bars.
\\
However, in order to draw more definitive conclusions and give more stringent
constraints much larger high-$z$ GRB complete samples, currently not
available in the literature, are required.


\section*{acknowledgments}

We acknowledge the anonymous referee for the extremely swift and constructive 
comments, and for all the useful bibliographic suggestions.
UM acknowledges invaluable technical support from the computing center of the
Max Planck Society, in Garching bei M\"unchen (Rechenzentrum Garching, RZG),
and kind hospitality at the Italian computing center (CINECA).
UM also acknowledges financial contribution from the Project ``HPC-Europa2'', 
grant number 228398, with the support of the European Community, 
under the FP7 Research Infrastructure Programme.
LM acknowledges financial contributions from contracts ASI
I/016/07/0 COFIS, ASI-INAF I/023/05/0, ASI-INAF I/088/06/0,
ASI `EUCLID-DUNE' I/064/08/0, PRIN MIUR 2009 ``Dark
energy and cosmology with large galaxy survey'', and PRIN INAF
2009 ``Towards an Italian network of computational cosmology''.
The bibliographic research for this work was done with the tools offered by 
the NASA Astrophysics Data System.
\\
\indent
UM wishes to dedicate this work to the memory of Paolo Borsellino
(1940 January 19 -- 1992 July 19), on the occasion of the twentieth 
anniversary of his death. ``Chi ha paura muore tutti i giorni'' (PB).


\bibliographystyle{mn2e}
\bibliography{bibl.bib}

\label{lastpage}
\end{document}